\documentclass[prb,twocolumn,superscriptaddress,showpacs,epsf]{revtex4}

\usepackage{graphicx}
\usepackage{latexsym}
\usepackage{amssymb}
\usepackage{amsfonts}
\usepackage{soul}
\usepackage{color}
\begin{document}
\bibliographystyle{apsrev}

\title{Effect of ferromagnetic film thickness on magnetoresistance
of thin--film superconductor--ferromagnet hybrids}

\author{A.Yu. Aladyshkin}
\email{aladyshkin@ipm.sci-nnov.ru} \affiliation{INPAC -- Institute
for Nanoscale Physics and Chemistry, Nanoscale Superconductivity
and Magnetism
\\ and Pulsed Fields Group, K.U.~Leuven, Celestijnenlaan 200D,
B--3001 Leuven, Belgium} \affiliation{Institute for Physics of
Microstructures, Russian Academy of Sciences, 603950 Nizhny
Novgorod, GSP-105, Russia}
\author{A.P. Volodin}
\affiliation{INPAC -- Institute for Nanoscale Physics and
Chemistry, Scanning Probe Microscopy Group, K.U.~Leuven,
Celestijnenlaan 200D, B--3001 Leuven, Belgium}
\author{V.V.~Moshchalkov}
\affiliation{INPAC -- Institute for Nanoscale Physics and
Chemistry, Nanoscale Superconductivity and Magnetism \\ and Pulsed
Fields Group, K.U.~Leuven, Celestijnenlaan 200D, B--3001 Leuven,
Belgium}


\date{\today}
\begin{abstract}
We study the influence of the thickness $D_f$ of the plain
ferromagnetic (F) film on the electrical resistance of the
flux-coupled hybrids, consisting of superconducting (S) Al film
and multilayer [Co/Pt] F film with out-of-plain magnetization. The
behavior of such hybrids at high and low temperatures is found to
be different: the nucleation of superconductivity at high
temperatures is governed mainly by the typical lateral dimensions
of the magnetic domains, while low temperature properties are
determined by topology of the magnetic template. We show that an
increase in the $D_f$ value leads to a broadening of the field-
and temperature intervals where non-monotonous dependence of the
superconducting critical temperature $T_c$ on the applied magnetic
field $H$ is observed (for demagnetized F films). Further increase
in the $D_f$ value results in a global suppression of
superconductivity. Thus, we determined an optimal thickness, when
the non-monotonous dependence $T_c(H)$ can be observed in rather
broad $T$ and $H$ range, what can be interesting for further
studies of the localized superconductivity in planar Al-based S/F
hybrids and for development of the devices which can exploit the
localized superconductivity.
\end{abstract}



\maketitle

\section{Introduction}

Recent technological achievements make it possible to fabricate
hybrid structures superconductor--ferromagnet (S/F) in which the
magnetic field produced by ferromagnetic elements varies in space
at submicron scales~\cite{Martin-JMMM-03} comparable with the
magnetic field penetration depth $\lambda$, the effective
penetration depth $\lambda^2/D_s$ or the superconducting coherence
length $\xi$ ($D_s$ is the thickness of superconducting film). In
the magnetically coupled S/F hybrids the interaction between the
superconducting and ferromagnetic subsystems occurs mainly via
slowly decaying stray field. It is natural to expect that the
magnetic field induced by the ferromagnet will strongly affect the
thermodynamical, magnetic and transport properties of the
considered hybrid systems (for review see Refs.
\cite{Martin-JMMM-03,Lyuksyutov-AdvPhys-05,Velez-JMMM-08,Aladyshkin-SuST-09}
and references therein). The properties of the ferromagnetic
superconductors and the S/F hybrids with rather strong exchange
interaction between superconductor and ferromagnet were discussed
in the reviews
\cite{Bulaevskii-AvdPhys-85,Izyumov-UFN-02,Buzdin-RMP-05,Bergeret-RMP-05}.
Hereafter we will focus only on the flux-coupled S/F hybrids.

It is already well known
\cite{Otani-JMMM-93,Aladyshkin-JPCM-03,Buzdin-PRB-03,Aladyshkin-PRB-03,Aladyshkin-PRB-06,Lange-PRL-03,Gillijns-PRB-06,Gillijns-PRB-07,Aladyshkin-PhysC-08,Gillijns-PRL-05,Yang-NatMat-04}
that the nonuniform magnetic field can modify the conditions for
the appearance of superconductivity due to the effect of a local
field compensation. This leads to an exotic non-monotonous
dependence of the superconducting critical temperature $T_c$ on an
external magnetic field $H$ applied perpendicular to the
superconducting film. Depending on the topology of the magnetic
field, nucleation of superconductivity in the compensated regions
(near the $|B_z|$ minima, $B_z$ is the perpendicular component of
the total magnetic field) results in either the field-induced
superconductivity, or the domain-wall superconductivity (DWS) and
the reverse-domain superconductivity (RDS). The field-induced
superconductivity is inherent for the S/F hybris with the arrays
of ferromagnetic dots,
\cite{Lange-PRL-03,Gillijns-PRB-06,Gillijns-PRB-07,Aladyshkin-PhysC-08}
whereas DWS and RDS are typical for the planar S/F hybrids with
domain structure in ferromagnetic
layer.\cite{Gillijns-PRB-07,Aladyshkin-PhysC-08,Gillijns-PRL-05,Yang-NatMat-04,Yang-PRB-06,Yang-ARL-06,Fritzsche-PRL-06,Aladyshkin-APL-09}
The problem of the formation of the localized superconductivity in
the presence of one-dimensional domain structure with out-of-plane
magnetization was considered theoretically in Refs.
\cite{Buzdin-PRB-03,Aladyshkin-PRB-03,Aladyshkin-PRB-06} It was
demonstrated that superconductivity localized near the magnetic
domain walls (DWS) at $H=0$ can be realized under certain
restrictions on the main parameters of the S/F hybrid,
\cite{Aladyshkin-PRB-06} namely: (i) the magnetization $M_0$ of
the ferromagnet, (ii) the width of the domains $L$, as well as
(iii) the thicknesses of the superconducting and ferromagnetic
films $D_s$ and $D_f$. On the contrary, the RDS regime in which
superconductivity appears above the magnetic domains of the
opposite polarity with respect to the $H$ sign, is considerably
less sensitive to these parameters. The RDS regime is always
realized for rather large $|H|$ values.\cite{Aladyshkin-SuST-09}
The influence of the non-uniform magnetic field of the magnetic
domains on the transport properties of the thin superconducting
films was recently studied for the following planar low--$T_c$ S/F
hybrid structures: \mbox{Nb/BaFe$_{12}$O$_{19}$}
(Refs.~\cite{Yang-NatMat-04,Yang-PRB-06}),
\mbox{Pb/BaFe$_{12}$O$_{19}$} (Ref.~\cite{Yang-ARL-06}),
\mbox{Nb/PbFe$_{12}$O$_{19}$} (Ref.~\cite{Fritzsche-PRL-06}),
\mbox{Al/BaFe$_{12}$O$_{19}$}
(Refs.~\cite{Aladyshkin-APL-09,Fritzsche-PRB-09,Aladyshkin-PhysicaC-10}),
\mbox{Al/[Co/Pt]$_n$}
(Refs.~\cite{Gillijns-PRB-07,Aladyshkin-PhysC-08}),
\mbox{Nb/[Co/Pt]$_n$} (Refs.~\cite{Gillijns-PRL-05,Zhu-PRL-08}),
\mbox{NbN/[Co/Pt]$_n$}
(Refs.~\cite{Rakshit-PRB-08a,Rakshit-PRB-08b}), \mbox{MoGe/GdNi}
(Ref.~\cite{Bell-PRB-06}), \mbox{Pb/FeNi}
(Ref.~\cite{VlaskoVlasov-PRB-08b}), \mbox{NbSe$_2$/FeNi}
(Ref.~\cite{VlaskoVlasov-PRB-08a}), \mbox{MoGe/FeNi}
(Refs.~\cite{Belkin-APL-08,Belkin-PRB-08}).

The main focus of the paper is to investigate experimentally the
influence of the thickness $D_f$ of the ferromagnetic film on the
electrical resistance of the thin-film planar S/F hybrids,
composed of superconducting Al film and multilayered [Co/Pt]$_n$
ferromagnetic film. Similar [Co/Pt]-based hybrid structures,
characterized by out-of-plane magnetic anisotropy, were recently
used for the studying of the influence of the inhomogeneous field
and the width of the domains on the shape of the phase transition
line $T_c(H)$.
\cite{Gillijns-PRL-05,Gillijns-PRB-07,Aladyshkin-PhysC-08} However
now we study the variation of the resistance not only at
temperatures close to the superconducting critical temperature
$T_c$, but also at low temperatures when the superconducting order
parameter is well developed.

We expect that the thickness of ferromagnet as well as the typical
width of the magnetic domain is also of importance for the
reentrant superconductivity. Indeed, the spatial distribution of
the magnetic field induced by the domain structure depends
substantially on the ratio between $D_f$ and $L$: the magnetic
field is almost uniform inside the domains at $L\ll D_f$ and it is
highly inhomogeneous in the opposite
limit.\cite{Aladyshkin-PRB-06} Thus, one can anticipate that the
thicker ferromagnetic films will have a stronger influence of the
superconducting properties of the S/F hybrid. The advantage of the
use of the ferromagnetic [Co/Pt]$_n$ multilayers is that we can
precisely control the thickness by choosing the number $n$ of the
bilayers in the preparation process.  We use thin aluminum films
as a superconducting material since aluminum has the maximal
coherence length (or the minimal upper critical field) among other
superconducting materials.\cite{Gillijns-PRB-07} Therefore thin Al
films should be very sensitive to the parameters of the built-in
magnetic field in the S/F hybrid.

We expect that the presented comparative study might be useful for
further investigations of the domain-wall superconductivity in the
S/F hybrids by means of local probe techniques (e.g., scanning
tunnelling microscopy/spectroscopy). Since at low temperatures the
scan range of the scanning tunnelling microscope (STM) is of the
order of $\mu$m, it is necessary to find the S/F system with
persistent small-scale domain structure and rather high amplitude
of the magnetic field. Ferromagnetic single crystals
BaFe$_{12}$O$_{19}$ and PbFe$_{12}$O$_{19}$ can be hardly used for
this purpose since the magnetic domains are rather large (from 2
to 30 $\mu$m) and they are strongly influenced by the external
magnetic field.\cite{Yang-NatMat-04,Fritzsche-PRL-06} On the other
hand, parallel magnetic domains in permalloy (FeNi) films have
sub-micron width, however they generate too weak stray field in
order to induce the localized superconductivity at low
temperatures.~\cite{Belkin-APL-08,Belkin-PRB-08} In this sense the
multilayered [Co/Pt] structures seem to be the most suitable for
the direct visualization of the DWS and the only question is what
are the optimal composition of such class of the S/F hybrids.
Finally, we believe that our study might be useful for development
of the superconducting electronic devices based on planar S/F
hybrid structures which can exploit the effect of the localized
superconductivity.

\section{Magnetic properties of ferromagnetic substrates}
\label{Section-MagneticProperties}

{\it Sample preparation.} In order to investigate the effects of
nonuniform magnetic field, induced by bubble magnetic domains, on
the magnetoresistance of thin superconducting Al films, three
planar hybrid samples were fabricated by means of molecular beam
epitaxy on SiO$_2$ substrates. The ferromagnetic part of the S/F
hybrids consists of a Pt(2.5~nm) buffer layer covered by a
multilayer of \mbox{[Co(0.4~nm)/Pt(1.0~nm)]$_{n}$}, where the
number of the bilayers $n$ was equal to 10, 15 and 20. All these
templates were covered with a 4 nm thick Si layer followed by a
superconducting Al layer of 20 nm thickness. Finally, the capping
Si layer (12 nm) was evaporated in order to prevent oxidation and
mechanical damage of the samples. It is important to notice that
all insulating/superconducting materials were evaporated
simultaneously in a single run, therefore the only difference in
the sample's composition is the thickness $D_f$ of the
ferromagnetic films: $D_f=16.5$~nm ($n=10$), $D_f=23.5$~nm
($n=15$), and $D_f=30.5$~nm ($n=20$). For brevity, we introduced
notations n--10, n--15 and n--20 in order to refer to the hybrid
samples containing the ferromagnetic films with 10, 15 and 20
bilayers, correspondingly. Since the Al films were electrically
insulated from the ferromagnetic substrate, we expect that the
interaction between ferromagnetic and superconducting parts in our
samples is purely magnetostatic in origin and the proximity effect
is negligible.

    \begin{figure}[t!]
    \begin{center}
    \includegraphics[width=8.5cm]{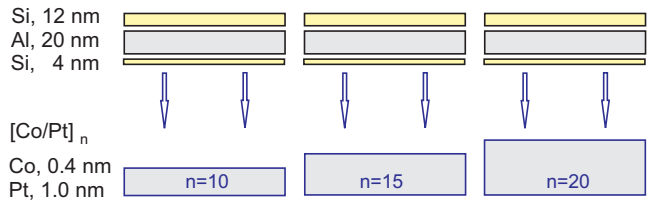}
    \end{center}
    \caption{(color online) Schematic presentation of the planar S/F hybrids
    under investigation.} \label{Fig-System}
    \end{figure}

    \begin{figure*}[hbt!]
    \begin{center}
    \includegraphics[width=0.95\textwidth]{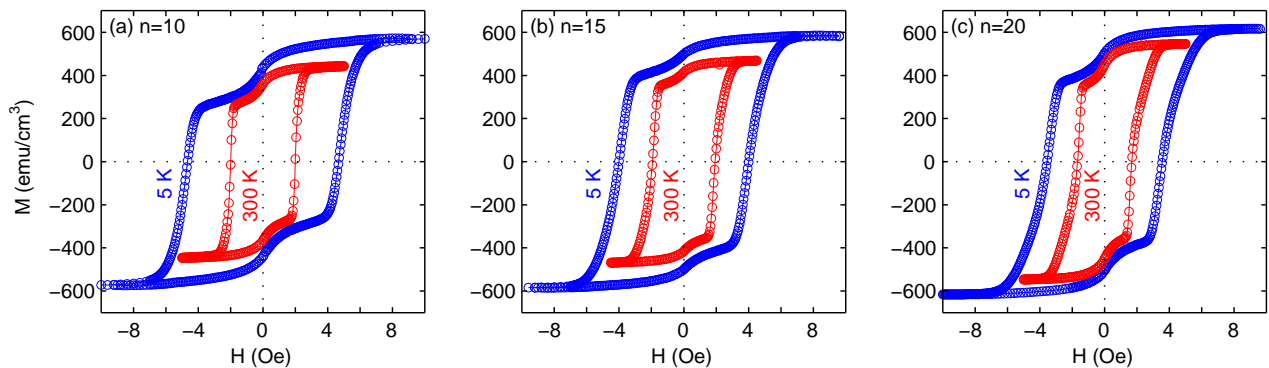}
    \end{center}
    \caption{(color online) Magnetization curves $M(H)$ for the
    ferromagnetic substrates consisting of Pt(2.5 nm) layer covered by
    a multilayer of \mbox{[Co(0.4 nm)/Pt(1.0 nm)]$_{n}$}, where the
    number of the bilayers are equal to (a) $n=10$,  (b) $n=15$, and (c) $n=20$.
    The inner $M(H)$ loops marked by red circles were measured at
    room temperature, while the outer loops (blue circles) correspond
    to $T=5$~K.} \label{Fig-MH-new}
    \end{figure*}

    \begin{figure}[b!]
    \begin{center}
    \includegraphics[width=0.35\textwidth]{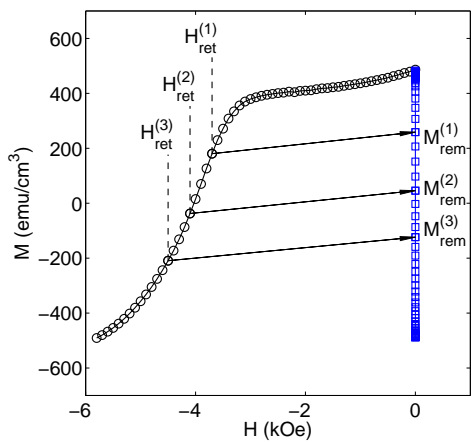}
    \end{center}
    \caption{(color online) Presentation of a method of preparation of the magnetic
    state with desirable remanent magnetization for a positively
    magnetized ferromagnetic film. The left curve (black circles) shows the
    descending branch of the typical $M(H)$ dependence measured for the
    sample n--15 at $T=5$~K. Depending on the returning field
    $H_{ret}<0$, one can get the remanent magnetization $M_{ret}$ at
    will (blue squares).} \label{Fig-Demagnetization}
    \end{figure}

{\it Magnetization loops $M$ vs. $H$.} The hysteresis $M(H)$ loops
of the ferromagnetic multilayered Co/Pt films of the different
thickness were measured using a commercial SQUID magnetometer in a
perpendicular applied field $H$ at $T=300$~K and $T=5$~K (above
the critical temperature of superconducting transition
$T_{c0}=1.45~$K), see Fig.~\ref{Fig-MH-new}. Such ferromagnetic
structures are known to possess well-defined out-of-plane
magnetization.\cite{Zeper-JAP-89} We demonstrated that the
increase in the thickness of the ferromagnetic film apparently
leads to a lowering of the coercive fields:
$H_{c,10}^{5K}=4.66$~kOe (n--10), $H_{c,15}^{5K}=4.02$~kOe
(n--15), and $H_{c,20}^{5K}=3.56$~kOe (n--20). We would like to
note that the saturated magnetization $M_s$ (the total magnetic
moment divided by the volume) at low temperatures is practically
independent on the thickness of the ferromagnetic film:
$M_{s,10}^{5K}=570$~emu/cm$^3$, $M_{s,10}^{5K}=580$~emu/cm$^3$,
$M_{s,10}^{5K}=610$~emu/cm$^3$. It is easy to see that the
increase in temperature substantially reduces the width of the
hysteresis loops (the corresponding coercive fields at $T=300$~K
are equal to $H_{c,10}^{300K}=1.99$~kOe,
$H_{c,15}^{300K}=1.93$~kOe and $H_{c,20}^{5K}=1.68$~kOe) and
decreases the saturated magnetization $M_s$.


    \begin{figure*}[t!]
    \begin{center}
    \includegraphics[width=0.95\textwidth]{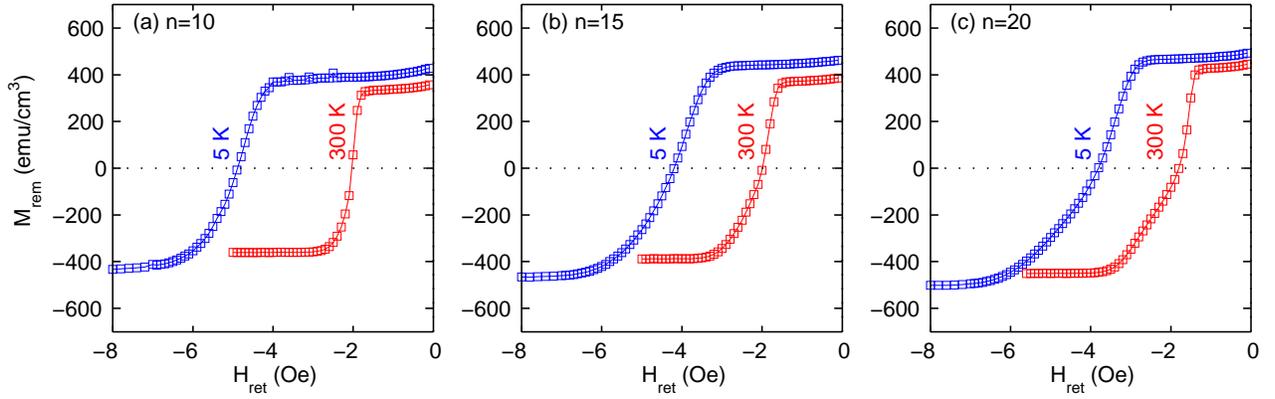}
    \end{center}
    \caption{(color online) Dependence of the remanent magnetization
    $M_{rem}$ of the ferromagnetic film \mbox{[Co(0.4 nm)/Pt(1.0 nm)]$_{n}$} on the returning field
    $H_{ret}$ measured at $T=300$~K and $T=5$: (a) sample n--10, (b) n--15, and (c) n--20.
    Note that all these curves
    $M_{rem}(H_{ret})$ are similar to the descending branches of the
    $M(H)$ loops but not identical to that since the remanent
    magnetization was always measured at $H=0$. The arrows mark the
    corresponding coercive fields $H_c$, where $M=0$.} \label{Fig-Remanent}
    \end{figure*}

Following to the idea proposed in Refs.
\cite{Gillijns-PRB-07,Aladyshkin-PhysC-08,Lange-APL-02}, we can
obtain any desirable remanent magnetization $M_{rem}$ of the
ferromagnet using the following procedure of incomplete
demagnetization: $H=0\Rightarrow 10$~kOe $\Rightarrow
H_{ret}\Rightarrow 0$ (Fig.~\ref{Fig-Demagnetization}), where
$H_{ret}$ is the so-called returning value, $H_{ret}<0$.  The
dependence $M_{rem}$ vs. $H_{ret}$ at $T=300$~K and $T=5$~K for
all samples are shown in Fig.~\ref{Fig-Remanent}. Although the
magnetic patterns, obtained by this procedure, appear to be
metastable, the prepared magnetic states are rather robust. At
least, as we can conclude from the reproducibility of experimental
data, they have not been substantially changed by the magnetic
field applied during the transport measurements (Section III).

    \begin{figure*}[t!]
    \begin{center}
    \includegraphics[width=13.0cm]{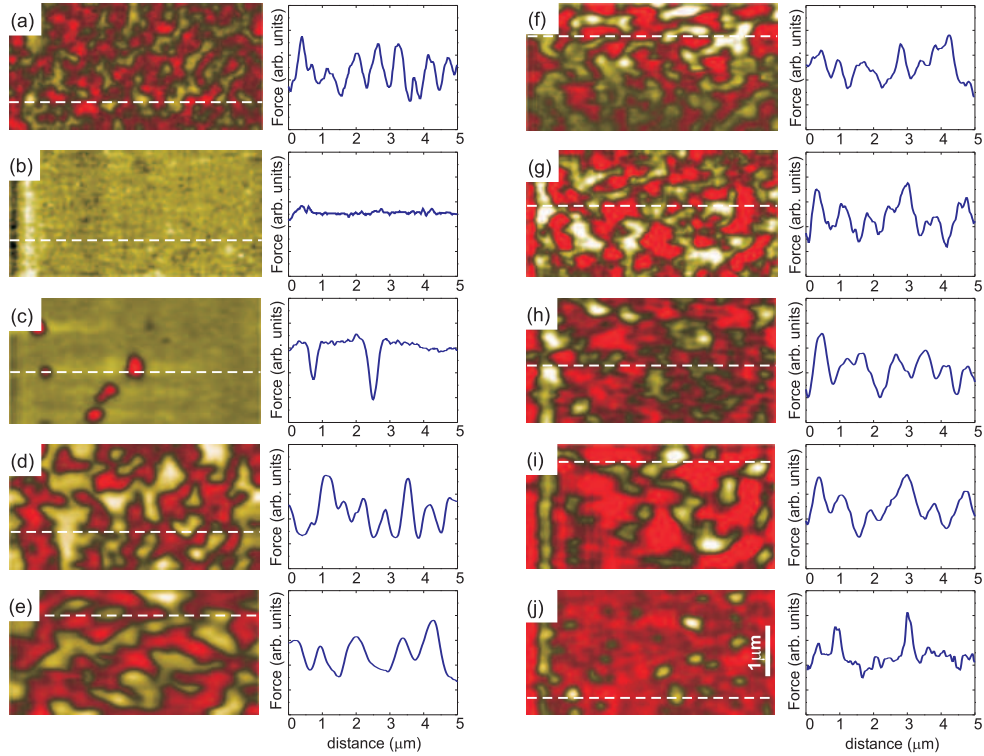}
    \end{center}
    \caption{(color online) Series of typical magnetic force
    microscopy (MFM) images for the sample n--15 measured
    at room temperature at $H=0$ after applying of the magnetic field
    $H=H_{ret}$ perpendicular to the film plane. The images show the
    the evolution of the magnetic domain structure over the same area
    5$\times$2.5~$\mu$m$^2$:
    (a) as-grown state,
    (b) positively magnetized state,
    (c) partially demagnetized state,
        $H_{ret}=-770$~Oe,  $|H_{ret}|/H^{300K}_{c,15}=0.40$,
    (d) $H_{ret}=-1060$~Oe, $|H_{ret}|/H^{300K}_{c,15}=0.55$,
    (e) $H_{ret}=-1330$~Oe, $|H_{ret}|/H^{300K}_{c,15}=0.69$,
    (f) $H_{ret}=-1710$~Oe, $|H_{ret}|/H^{300K}_{c,15}=1.00$,
    (g) $H_{ret}=-1930$~Oe, $|H_{ret}|/H^{300K}_{c,15}=1.00$,
    (h) $H_{ret}=-2130$~Oe, $|H_{ret}|/H^{300K}_{c,15}=1.10$,
    (i) $H_{ret}=-2220$~Oe, $|H_{ret}|/H^{300K}_{c,15}=1.15$,
    (j) $H_{ret}=-2390$~Oe, $|H_{ret}|/H^{300K}_{c,15}=1.24$
    The right-hand side panels of the images show normalized line traces
    of the measured magnetostatic forces taken across the the dashed
    lines in the correspondent images.}
    \label{Fig-MFM}
    \end{figure*}

{\it MFM study of magnetic patterns.} An almost perfect
coincidence of the normalized hysteresis loops $M(H/H_c)/M_s$ for
$T=300$~K and $T=5$~K indicates that the magnetization reversal
processes at low and room temperatures are expected to be quite
similar. Therefore magnetic force microscopy (MFM) performed at
room temperature can provide an additional valuable information
concerning a reshaping of the bubble magnetic domains in the
ferromagnetic films at low temperatures, occurring during the
process of the incomplete demagnetization.

    \begin{figure*}[htb!]
    \begin{center}
    \includegraphics[width=0.95\textwidth]{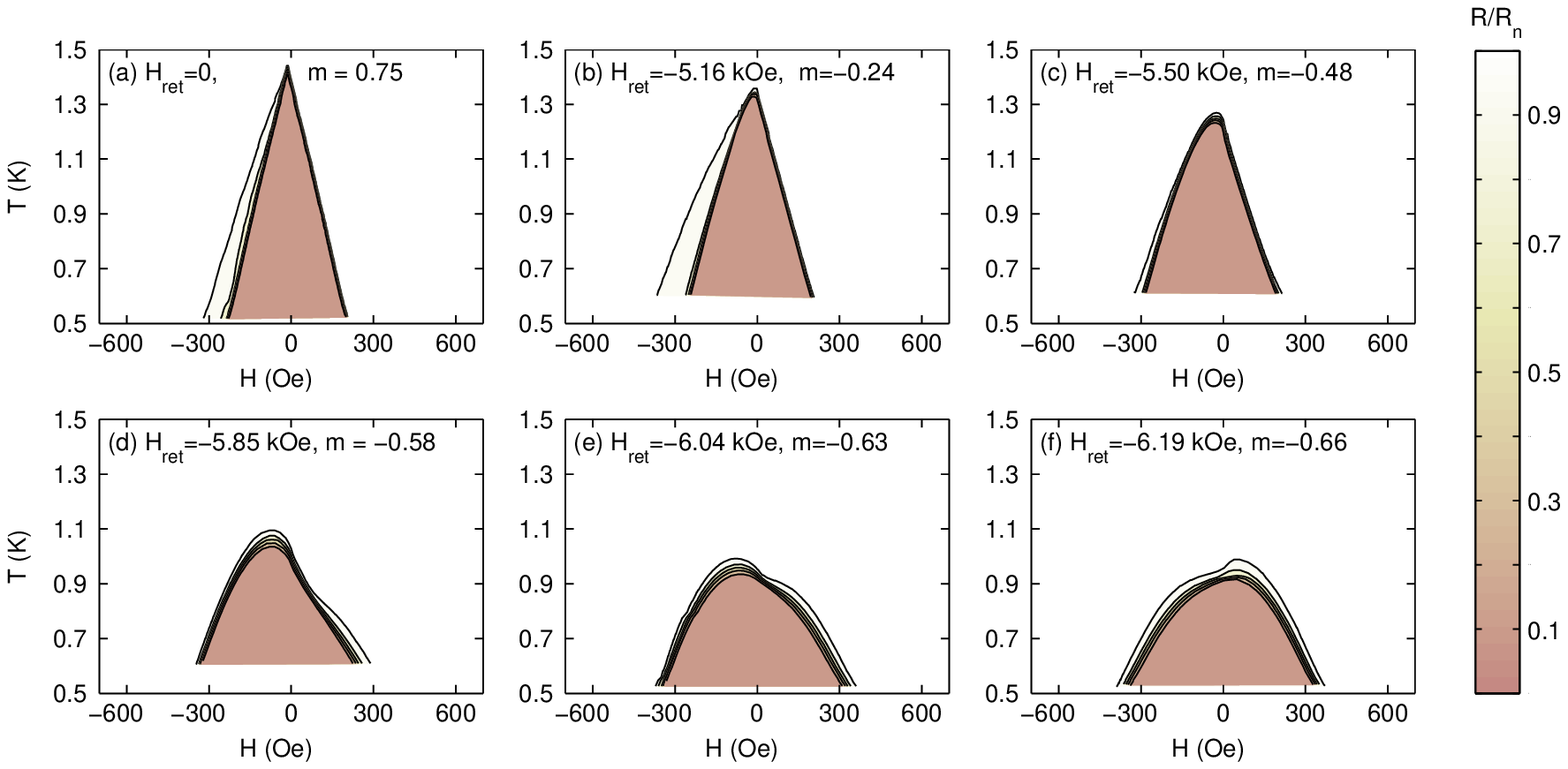}
    \end{center}
    \caption{(color online) Dc resistance of the sample n--10 as a function of $H$ and $T$ in different magnetic
    states. The returning $H_{ret}$ fields and the normalized remanent magnetization $m=M_{ret}/M_{s,10}^{5K}$ are indicated on the plots.
    The darker shades correspond to lower resistance values. The
    solid lines are the level curves $R(H,T)=\alpha R_n$, where $R_n$
    is the resistance in normal state, $\alpha=0.1,\ 0.3,\, 0.5,\,
    0.7$ and 0.9.} \label{Fig-Summary-20-10}
    \end{figure*}

The magnetic domain structure corresponding to the different
remanent states in the Co/Pt multilayers was investigated by MFM
technique at room temperature in the presence of the external
magnetic field similar to that reported in
Ref.~\cite{Yang-NatMat-04} The MFM measurements were carried out
using a commercial Autoprobe M5 (Veeco Instruments) scanning probe
microscope. The MFM tip was maintained at a fixed ``flying" height
(10--20 nm) above the sample in order to exclude the tip influence
effects which can modify the magnetic domain structure. The tip
oscillation amplitude was about 20 nm. With indicated scan heights
in the flying mode this equates to an average tip-to-sample
distances of 30--40 nm. The standard non-contact mode of scanning
probe microscopy with ac cantilever amplitude stabilization was
used. The dc deflection response of the MFM cantilever was used to
plot the magnetic interactions between the tip and sample as a
function of the tip position. The magnetic force interaction is
related to the vertical component of the spatial derivative of the
magnetic field from the sample. Therefore, MFM is sensitive to the
strength and polarity of near-surface stray fields produced by
ferromagnetic samples. This allows to obtain information about the
overall domain topology from an MFM image. A variable magnetic
field module providing uniform fields up to 4 kOe in the direction
perpendicular to the sample surface was mounted on the MFM sample
stage to study the evolution of the domain structure with varying
$H$. The magnetic field module was used also to
magnetize/demagnetize of the samples in-situ.

Figure \ref{Fig-MFM} summarizes the evolution of the magnetic
domains obtained for the sample n--15 for the different returning
field values. We refer to this sample as the most representative.
This figure shows the MFM images of the {\it same} area acquired
at different magnetization states of the sample. In the first
approximation, light and dark regions can be attributed to
magnetic domains with different directions (signs) of the normal
component of the local magnetic field, respectively. The
right-hand side panels of the images show the normalized line
traces of the measured magnetostatic forces taken across the the
dashed lines in the correspondent images.

The panel (a) in Fig.~\ref{Fig-MFM} shows the domain pattern for
the Co/Pt film in the as-grown state (before applying any magnetic
field). Clearly, the length scales of the magnetic field
variations appear to be smallest (the averaged width is about
0.3~$\mu$m) and comparable with superconducting coherence length
(0.1~$\mu$m at $T=0$, see estimates below). By magnetizing the
sample, we remove all magnetic domains of the opposite direction
with respect to the direction of the external magnetic field, and
therefore there is no noticeable MFM contrast in the
Fig.~\ref{Fig-MFM}(b). Then, increasing the applied field, one
gradually changes the shape and the size the magnetic domains of
both polarities [panels (c)--(f) in Fig.~\ref{Fig-MFM}]. The
appearance of the isolated negative domains in the positively
magnetized matrix in the presence of the external magnetic field
can be a plausible cause of the plateau on the descending branches
of the M(H) dependencies [see panels (a)-(c) in Fig.
\ref{Fig-MH-new}].  The size of the magnetic domains appears to be
the smallest (but still larger than that in the as-grown state) at
$|H|\simeq H_c$ [see Fig.~\ref{Fig-MFM} (g)]. Figure clearly
portrays that the widths of positive and negative domains are
rather close. This observation is consistent with the results
presented in Fig. \ref{Fig-Remanent} which demonstrates that the
remanent magnetization is close to zero. The final stage of the
magnetization reversal is typical for ferromagnetic [Co/Pt] films:
negative domains continue to grow and become dominant while the
positive domains shrink in size [see Fig.~\ref{Fig-MFM} (h--j)].
The gradual increase of the area covered by the magnetized domains
at increasing $|H_{ret}|$, what corresponds to the observed
monotonous decrease of the magnetization, was confirmed by the
performed analysis of the domain coverage.

    \begin{figure*}[t!]
    \begin{center}
    \includegraphics[width=0.95\textwidth]{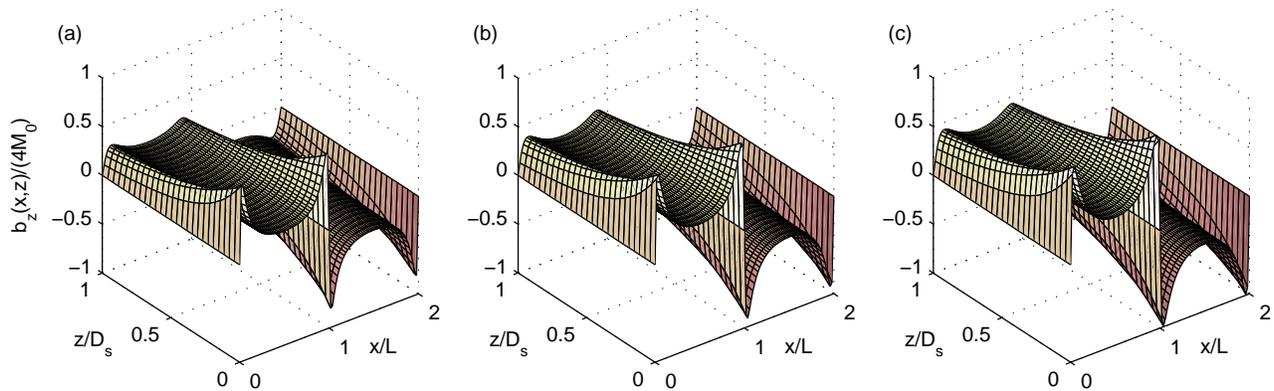}
    \end{center}
    \caption{(color online) The examples of the spatial distribution of the $z-$component of the magnetic field
    induced by one-dimensional periodic domain structure with out-of-plane
    magnetization inside the superconducting film for (a) $D_f=16.5$ nm, (b) $D_f=23.5$ nm, and (c) $D_f=30.5$ nm.
    The regions $-D_f<z<0$ and $0<z<D_s$ correspond
    to the ferromagnetic and superconducting films, respectively.
    All other relevant parameters ($D_s$, equilibrium half-period $L$,
    the spacer $h$ between the superconducting and ferromagnetic films, the magnetization $M_0$ etc.)
    are taken to be typical for our planar S/F structures,
    $x-$axis is taken perpendicular to the domain walls,
    see also Refs.~\cite{Gillijns-PRL-05,Gillijns-PRB-07}} \label{Fig-Field}
    \end{figure*}

\section{Transport properties of S/F hybrids}

The measurements of the dc electrical resistance $R$ as a function
of temperature $T$ and the external magnetic field $H$, applied
perpendicularly to the plane of the S/F hybrid structures, were
carried out in a commercial Oxford Instruments cryostat. The
resistance was measured in the large-area specimens by a standard
four-probe technique. The magnetic field produced by a
superconducting magnet (up to $10^4$ Oe) was used for a
magnetization/demagnetization of the sample in-situ before the
magnetoresistive measurements. The range of the $H$ sweeping (from
-700 Oe to 700 Oe) appears to be considerably smaller than the
corresponding coercive fields of the ferromagnetic films at low
temperatures (see Section II). Therefore we anticipate that the
parameters of the domain structure, prepared by incomplete
demagnetization, remains almost unaltered.

\subsection{10 bilayer sample}

We begin with the discussion about the transport properties of the
sample n--10 with the thinnest ferromagnetic [Co/Pt]$_{10}$
substrate.

    \begin{figure*}[htb!]
    \begin{center}
    \includegraphics[width=0.95\textwidth]{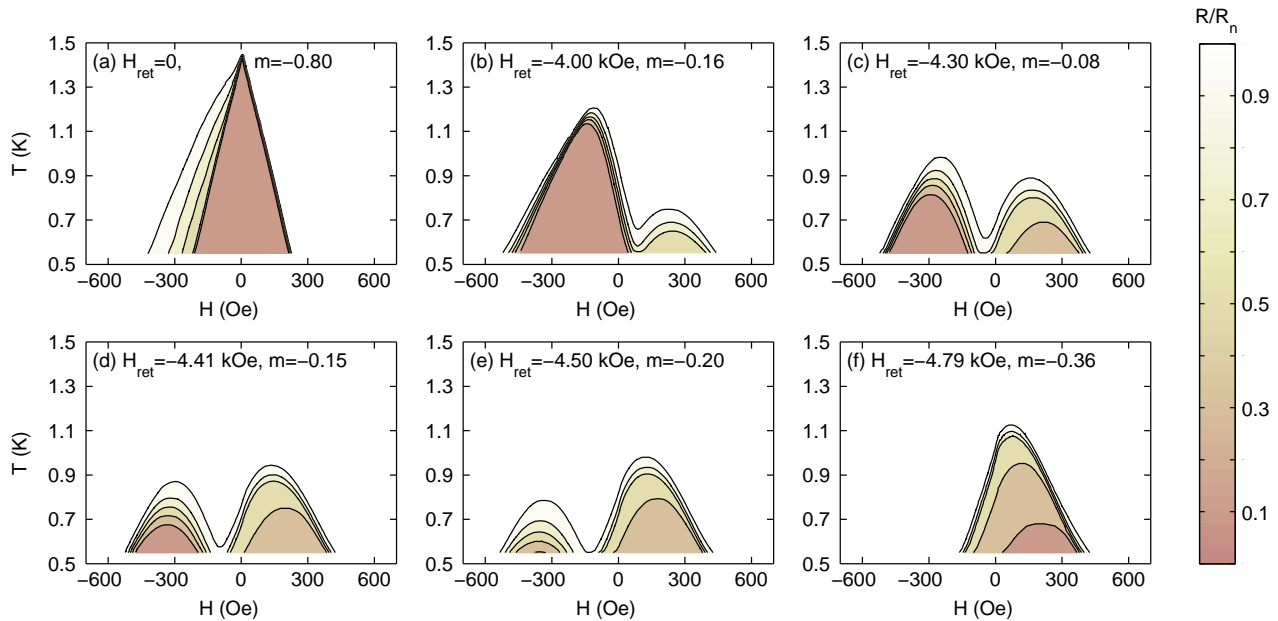}
    \end{center}
    \caption{(color online) Dc resistance of the sample n--15 as a function of $H$ and $T$ in different magnetic
    states. The returning $H_{ret}$ fields and the normalized remanent magnetization $m=M_{ret}/M_{s,15}^{5K}$ are indicated on the plots.
    The darker shades correspond to lower dc resistance
    values. The solid lines are the level curves $R(H,T)=\alpha R_n$,
    where $R_n$ is the resistance in normal state, $\alpha=0.1,\, 0.3,\,
    0.5,\, 0.7,\,$ and 0.9.} \label{Fig-Summary-20-15}
    \end{figure*}

For this sample which was preliminary magnetized in the positive
direction we found out that the all level curves of the resistance
$R(H,T)=\alpha R_n$ have the maxima positioned almost at $H=0$,
where $R_n$ is the normal state resistance [panel (a) in
Fig.~\ref{Fig-Summary-20-10}]. It is worth noting that two level
curves for $\alpha=0.7$ and 0.9) are asymmetrical with respect to
$H=0$. In the other words, for a given temperature the resistance
of the sample starts to deviate from its normal value at larger
$|H|$ values for $H<0$ than for $H>0$. Currently we have no
reliable interpretation of this asymmetry. A possible explanation
is that the negative magnetic field $|H|>50$~Oe induces the
negatively magnetized domains, which disappear at the $H$
inversion. However in our room temperature MFM measurements in the
presence of the external magnetic field $H=-200$~Oe and
$H=+200$~Oe we did not observe any change of the magnetic patterns
and an appearance of such magnetic domains of the opposite
polarity. We can attribute this to the restricted scan range and
insufficient sensitivity of our MFM microscope.

Nevertheless, the other level curves corresponding to low
resistance criteria ($\alpha=0.1, 0.3$ and 0.5) are almost
symmetrical, and they practically coincide. It means that the
conditions for a dropping of the resistance to zero are
polarity-independent. Indeed, even if the isolated magnetic
domains of opposite polarity appear reversibly at $H<0$, they
cannot ensure a vanishing of the electrical resistance. It is
quite natural to attribute this lines to the upper critical field
expression $H_{c2}=H^{(0)}_{c2}\left(1-T/T_{c0}\right)$, where the
maximal critical temperature $T_{c0}=1.45$~K and the upper
critical field $H^{(0)}_{c2}=368$~Oe at $T=0$ were found from the
optimal fitting. We would like to mention that the fitting gives
us the same parameters $T_{c0}$ and $H^{(0)}_{c2}$ for all three
hybrid samples as we demonstrate below. This can be considered as
an evidence of that the Al films have identical superconducting
characteristics. Using the $H^{(0)}_{c2}$ value, we also estimated
of the superconducting coherence length
\mbox{$\xi_0=\sqrt{\Phi_0/(2\pi H^{(0)}_{c2})}$}=93~nm at $T=0$.

During the procedure of the preparation of the magnetic state, one
can irreversibly create negative magnetic domains by applying the
external field $H_{ret}$ of the order of the coercive field
$H^{5K}_{c,10}$, as it follows from the MFM measurements at room
temperature. The reduction of the averaged remanent magnetization
results in the substantial suppression of the $T_c$ maximum (up to
50\%) [panels (b), (c) and (d) in Fig.~\ref{Fig-Summary-20-10}].
The similar depletion of the maximal critical temperature was
obtained for the same sample in as-grown state (not presented in
this paper). At the same time the all maxima in the level curves
$R(H,T)=\alpha R_n$ are shifted to negative $H$ values of the
order of 50--80 Oe depending on the $H_{ret}$ values. This finding
can be explained by a field compensation effect above the domains
with positive magnetization which are responsible for a
percolation of the supercurrents at the initial stage of the
magnetization reversal. Even a tiny change of the returning field
essentially modifies the magnetoresistance of the hybrid
structure: the maxima at the level curves which were initially
positioned at $H<0$, shift toward $H>0$ and vice versa [panels
(e)--(f) in Fig.~\ref{Fig-Summary-20-10}]. In other words, we
cannot observe the reentrant superconductivity for this sample for
any $H_{ret}$ parameter. In comparison with the samples n--15 and
n--20 (see below) the displacement of the $T_c$ maximum
($\sim$80~Oe) is too small. This indicates that both the amplitude
and typical length scales of the built-in magnetic field are not
large enough for a localizing of the order parameter in the
compensated regions.

\subsection{15 bilayer sample}

More interesting results were obtained for the sample with a 15
bilayer ferromagnetic film (the case of intermediate thickness of
ferromagnetic substrate). One might conclude that an increase in
the ferromagnetic film thickness will lead to

(i) an increase of the typical width of the magnetic domains,
since the equilibrium period of the domain structure with
out-of-plane magnetization is proportional to \mbox{$\sqrt{D_f}$},
Ref. \cite{Landau};

(ii) an increase of the absolute value of the flux produced by the
magnetic domains, since the smaller the $D_f$ value, the faster
the decay of the magnetic field in the direction, perpendicular to
the film plane, is;

(iii) a decrease of the width of the magnetization curve, as it
was revealed in the section \ref{Section-MagneticProperties}.

The first case (i) is difficult to verify in our experiments,
since the parameters of the prepared domain structure are much
more sensitive to the returning field values rather than to the
thickness of the substrate. As far as the second case (ii) is
concerned, we analyzed the spatial distribution of the
$z-$component of the magnetic field induced by one-dimensional
periodic domain structure, characterized by the period $L$ and the
magnetization $M_s$ (Fig.~\ref{Fig-Field}). It is obvious that the
increase in the thickness of ferromagnet raises the averaged field
(or magnetic flux) emanating from the domains, since the magnetic
field decays slower in the transverse direction as $D_f$ increases
(for a given domain's width).

The results of the transport measurements are summarized in
Fig.~\ref{Fig-Summary-20-15}. Analogous results obtained for the
similar hybrid structure Al(50~nm)/[Co/Pt]$_{15}$ were partly
presented in Refs. \cite{Gillijns-PRB-07,Aladyshkin-PhysC-08}, in
the form of a single curve of the constant resistance of the
sample. However here we present complete information concerning
the variations of magnetoresistance in full $T-H$ plane.

First of all we would like to note that the maximal critical
temperature $T_{c0}=1.45$~K of the sample n--15 in the fully
magnetized state is the same as for the samples n--10 and n--20.
In addition, all the level curves of the resistance for the
uniformly magnetized sample are coincident straight lines (for the
resistance ratios $\alpha=0.1$ and 0.3) with the almost same slope
$dT_c/dH\simeq T_{c0}/H^{(0)}_{c2}$ as for the samples n--10 and
n--20. It clearly indicates that there are no differences in the
superconducting properties of all hybrid samples.

    \begin{figure*}[htb!]
    \begin{center}
    \includegraphics[width=0.95\textwidth]{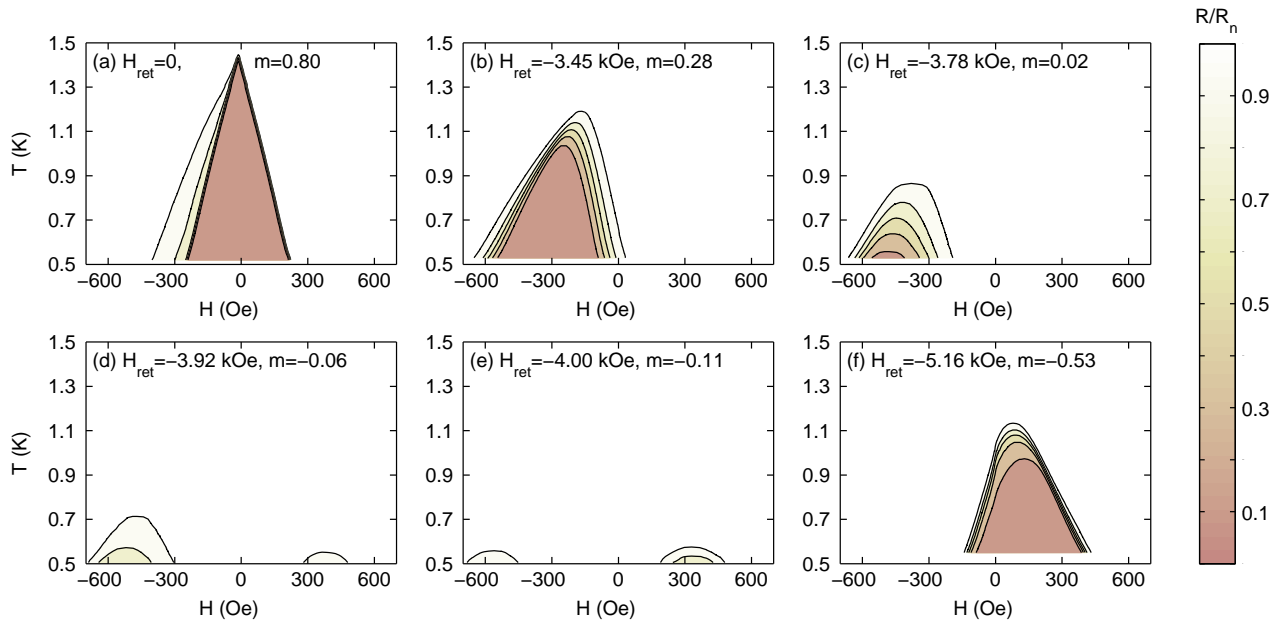}
    \end{center}
    \caption{(color online) Dc resistance of the sample n--20 as a function of $H$ and $T$ in different magnetic
    states. The returning $H_{ret}$ fields and the normalized remanent magnetization $m=M_{ret}/M_{s,20}^{5K}$ are indicated on the plots.
    The darker shades correspond to lower dc resistance
    values. The solid lines are the level curves $R(H,T)=\alpha R_n$,
    where $R_n$ is the resistance in normal state, $\alpha=0.1,\, 0.3,\,
    0.5,\, 0.7,\,$ and 0.9.} \label{Fig-Summary-20-20}
    \end{figure*}

The formation of isolated negative magnetic domains in the
ferromagnetic film which is preliminary magnetized in a positive
direction leads to a shift of the main $T_c$ maximum toward
negative $H$ values and to a suppression of its amplitude [panel
(b) in Fig.~\ref{Fig-Summary-20-15}]. This can be easily explained
by a compensation of the local magnetic field above rather wide
positive magnetic domains. Such compensation leads to a preferred
nucleation of superconductivity there \cite{Gillijns-PRB-07} (the
regime of the reverse-domain superconductivity). At the same time
the nucleation of superconductivity above negative magnetic
domains, which have smaller typical width than positive domains
(see Fig.~\ref{Fig-MFM}) is energetically unfavorable. As a
result, the $T_c$ maximum at $H>0$  has smaller amplitude. Indeed,
as we argued in Refs. \cite{Gillijns-PRB-07,Aladyshkin-PhysC-08},
the larger the area $\ell$ where the superconducting wave function
is confined, the higher the critical temperature $T_c$ of the
appearance of such localized superconducting state is:
$1-T_c/T_{c0}\simeq \xi_0^2/\ell^2$. This effect can be
interpreted as a quantum size effect for the localized
superconducting wave function in non-uniform magnetic field, what
is analogous to the standard quantum size effect for single
electron wave function in a potential
well.\cite{Aladyshkin-SuST-09} In addition, the maxima on all the
level curves are also shifted, pointing out that superconductivity
above positive domains is responsible for a percolation for
supercurrents at lower temperatures.

A further increase of the $H_{ret}$ value leading to a more
pronounced decrease of the size of the positively magnetized
domains, causes a suppression of the $T_c$ maximum at $H<0$ [panel
(c) in Fig.~\ref{Fig-Summary-20-15}]. This peak is now shifted to
even higher negative fields, since the absolute value of the
$z-$component of the stray field, (which has to be compensated),
increases as the typical domain width decreases. Simultaneously,
the growth of negatively magnetized domains results in a more
favorable order parameter nucleation above negative domains and,
accordingly, a second peak in the critical temperature at $H>0$
becomes higher. Thus, two peaks corresponding to the
reversed-domain superconductivity both above positive and negative
domains are easily seen in the $H-T$ diagram. Taking the positions
of these $T_c$ maxima, once can estimate the amplitude of the
nonuniform magnetic field, which is of the order of 300~Oe. In the
vicinity of the coercive field even small $H_{ret}$ variation can
induce a evident change in the relative amplitude of the $T_c$
peaks, which should be attributed to the reshuffling or reshaping
of the magnetic domains [compare panels (c) and (d) in
Fig.~\ref{Fig-Summary-20-15}]. We would like to note that the
level curves for high and low resistance ratio (e.g. $\alpha=0.1$
and $0.9$) may have completely different shapes. For instance, the
nucleation of superconductivity (i.e. an initial decrease of the
resistance) can occur both at the positive and negative
compensation fields almost at the same temperature, however the
sample's resistance goes to zero only at $H<0$ [panel (d) in
Fig.~\ref{Fig-Summary-20-15}]. We interpret this finding as a
consequence of the effect of current percolation in a irregular
labyrinth-type domain pattern.

Interestingly, the shape of the level curve for $\alpha=0.9$ quite
well resembles the non-monotonous dependence $T_c(H)$ with
positive curvature at small $H$ values, predicted theoretically
for the regime of the domain-wall
superconductivity.\cite{Buzdin-PRB-03,Aladyshkin-PRB-03,Aladyshkin-PRB-06}
We would like to interpret our result as a fingerprint of the
possible crossover between reverse-domain superconductivity and
domain-wall superconductivity at sweeping $H$ [panels (c)--(e) in
Fig.~\ref{Fig-Summary-20-15}].


For higher $H_{ret}$ values the first peak, located at negative
fields, disappears, whereas the peak at positive fields shifts up
in temperature and is displaced to a lower field curve [panels
(e)--(f) in Fig.~\ref{Fig-Summary-20-15}]. This second peak will
eventually evolve into a linear phase boundary when the
ferromagnetic film becomes fully magnetized in the negative
direction.

\subsection{20 bilayer sample}

Finally, we briefly discuss the superconducting properties of the
sample n--20 with the thickest ferromagnetic substrate. The
general evolution of the magnetoresistance at varying $H_{ret}$
parameter is quite similar to that for the sample n--15 (compare
Figs.~\ref{Fig-Summary-20-15} and \ref{Fig-Summary-20-20}). For
the sample in the demagnetized state ($|H_{ret}|\sim H_c$) the
effect of the stray field of the magnetic domains is so strong
that we can observe only the initial stage of the decrease of the
electrical resistance at the compensation fields in very narrow
range of available temperatures [panels (d) and (e) in
Fig.~\ref{Fig-Summary-20-20}]. This displacement of the $T_c$
maxima to the higher $H$ values (400--500 Oe) as well as the
lowering of the critical temperature at $H=0$ are in agreement
with our explanation concerning the increase of the averaged
magnetic field produced by the magnetic domains in thick
ferromagnetic films.

Currently we cannot explain the observed global suppression of the
maximal critical temperature near the compensation fields [panels
(d) and (e)]. For instance, this effect can be caused by
considerable shrinkage of the domain width down to the length
scales comparable with $\xi_0$. However, typical width of magnetic
domains for sample n--20 does not differ considerably from that
for the sample n--15 as it follows from the MFM measurements.
Therefore we might expect the essential modification of the
remanent distribution of magnetization for thick ferromagnetic
film, e.g. an increase in the in-plane component of the magnetic
moments. Anyway, the clarification of the origin of the global
suppression of superconductivity in the S/F hybrids with thick
multilayered ferromagnetic films deserves a separate detailed
study.

\section{Conclusion}

We investigated experimentally the peculiarities of the electrical
transport on large-area superconducting aluminum films deposited
on top of multilayered ferromagnetic [Co/Pt]$_n$ structure ($n$ is
the number of the bilayers). We demonstrated that: (i) by changing
the $n$ value during the fabrication process and (ii) by varying
the remanent magnetization of the hybrid sample during the
demagnetization procedure, one can control a spatial distribution
of the magnetic field, induced by magnetic domains, inside the
superconducting film. We showed that the nonuniform magnetic field
generated by magnetic bubble domains strongly modifies the
dependence of the electrical resistance of the sample on
temperature $T$ and the external magnetic field $H$. We found out
that the behavior of the planar S/F hybrids at temperatures close
to the critical temperature of the superconducting transition and
at low temperatures can be completely different: the nucleation of
superconductivity is governed mainly by the typical lateral
dimensions of the magnetic domains regardless their shapes, while
at low temperatures the general topology of the magnetic pattern
becomes very important.

We demonstrated that the increase in the ferromagnetic film
thickness leads to the more pronounced reentrant
superconductivity, characterized by a decrease of the resistance
as $H$ increases. However further increase of the $n$ number
suppresses the critical temperature of the hybrid sample. Thus for
the planar Al-based hybrid structures we determined an optimal
thickness of the ferromagnetic film ($n=15$), when the
non-monotonous dependence of the superconducting critical
temperature on $H$ can be observed in rather wide $T$ and $H$
range. This finding can be of interest, e.g., for further
experimental investigations of the different regimes of the
localized superconductivity in the presence of magnetic templates.
Indeed, the observed parabolic dependence $T_c=\alpha_0+\alpha_1
|H|^2$ can indicate the superconductivity localized near magnetic
domain walls ($\alpha_0$, $\alpha_1$ are positive constants).
Nevertheless the problem concerning the direct visualization of
the domain-wall superconductivity in the S/F hybrids is still
challenging. The formation of the localized superconductivity
(both the domain-wall superconductivity and the reverse-domain
superconductivity) can be detected by means of the scanning
tunnelling microscopy/spectroscopy (STM/STS). However a matching
between the magnetic pattern and the distribution of the local
density of states can be performed in a case of domain structure
with unalterable small-scale domain structure. In addition, rather
high amplitude of the magnetic field is necessary to guarantee the
localization of the superconducting wave function near the domain
walls. Thus, the parameters of the domain structure in the
multilayered [Co/Pt]$_n$ films meet these requirements. Therefore
the planar S/F hybrids with the optimal composition, resulting in
the well-defined reentrant superconductivity, can be effectively
used for the high-resolution STM/STS investigations of the systems
revealing the localized superconductivity.

We thank W. Gillijns and B. Opperdoes for technical support and D.
Roditchev for valuable comments. This work was supported by the
Methusalem Funding of the Flemish Government, the NES -- ESF
program, the Belgian IAP, the Fund for Scientific Research --
Flanders (F.W.O.--Vlaanderen), the Russian Fund for Basic
Research, RAS under the Program ``Quantum physics of condensed
matter", Russian Agency of Education under the Federal Program
``Scientific and educational personnel of innovative Russia in
2009--2013" and the Presidential grant MK-4880.2008.2.

\end{document}